\documentclass{pasj00}

\SetRunningHead{H. Hayashi and M. Chiba}{
Dynamical Effects of CDM Subhalos on a Galactic Disk}
\Received{}
\Accepted{}

\begin{document}

\title{Dynamical Effects of CDM Subhalos on a Galactic Disk}

\author{Hirohito \textsc{Hayashi} and Masashi \textsc{Chiba}}

\affil{Astronomical Institute, Tohoku University, Sendai 980-8578}

\email{hayashi@astr.tohoku.ac.jp, chiba@astr.tohoku.ac.jp}


\maketitle

\begin{abstract}
We investigate the dynamical interaction between a galactic disk and
surrounding numerous dark subhalos as expected for a galaxy-sized halo
in the cold dark matter (CDM) models. Our particular interest is to what
extent accretion events of subhalos into a disk are allowed in light of
the observed thinness of a disk. Several models of subhalos are considered
in terms of their internal density distribution, mass function, and spatial
and velocity distributions. Based on a series of N-body simulations,
we find that the disk thickening quantified by the change of its scale
height, $\Delta z_d$, depends strongly on the individual mass of an
interacting subhalo $M_\mathrm{sub}$. This is described by the relation,
$\Delta z_d/R_d \simeq 8 \sum_{j=1}^N (M_{\mathrm{sub},j}/M_d)^2$,
where $R_d$ is a disk scale length, $M_d$ is a disk mass,
and $N$ is the total number of accretion events of subhalos
inside a disk region ($\le 3R_d$).
Using this relation, we find that an observed thin disk has
not ever interacted with subhalos with the total mass of more than 15~\%
disk mass. Also, a less massive disk with smaller circular velocity
$V_c$ is more affected by subhalos than a disk with larger $V_c$,
in agreement with the observation. Further implications of our results
for the origin of a thick disk component are also discussed.
\end{abstract}

%%% Sec.1 %%%
\section{Introduction}

The cold dark matter (CDM) paradigm has become a standard framework for
understanding the structure formation in the Universe.
According to this theoretical paradigm, the growing process of
self-gravitating structures is hierarchical in the sense that small dark
matter halos virialize first, and aggregate successively into larger and
larger objects. This clustering process of dark matter halos
is successful for explaining a wide variety of observations including
the large-scale distribution of galaxies.

In this CDM scenario, N-body simulations are an important tool in order to
investigate the non-linear growth of cosmological structures.
Early N-body simulations based on the CDM models
suffered from the so-called {\it over-merging} problem, i.e., substructures
are disrupted very quickly within dense environments (\cite{summers1995}).
However, recent high-resolution N-body simulations have revealed
the presence of hundreds of dark matter substructures (subhalos) which
survive in not only cluster scales but also galactic scales (\cite{moore1999};
\cite{klypin1999}). This large number of subhalos in a galaxy-sized halo is in
contrast to only about a dozen satellite galaxies in the Galaxy, which
confronts so-called ``the Missing Satellite Problem''.
Several authors have argued that this apparent discrepancy could
be resolved by considering some suppressing process for star formation,
such as gas heating by an intergalactic ionizing background or energy
feedback from evolving stars. In whatever models relying on the suppression
of galaxy formation, a typical galaxy-sized halo
should contain numerous dark subhalos.

Then, there is a possibility that a large amount of subhalos interact
frequently with a stellar disk embedded in the center of a halo, so that
the disk would be dynamically heated and thickened.
On the other hand, an observed galactic disk is rather thin: the scale
height (or half thickness) is only about $\sim 250$ pc in the Galaxy.
Likewise, recent observations of external disk galaxies
(\cite{kregel2002}) suggest that the observed scale height of a disk,
$z_d$, is confined to some limiting value relative to the scale length
of a disk, $R_d$, i.e, $z_d / R_d < 0.2$.

This observed thinness of a disk provides important limits on the disk
heating due to infalling satellites. T\'{o}th \& Ostriker (1992) analytically
evaluated this effect and concluded that an observed disk like that of the
Galaxy within the solar radius should have interacted with satellites with
no more than 4~\% of the present disk mass within the last 5 Gyr.
Subsequent numerical simulations of an interaction between a disk and a single
satellite (e.g. \cite{velazquez1999}) showed that their analytical estimation
for the disk heating was somewhat too high because an actual interaction
process is highly non-linear and more complicated than simplified analytical
representation. Interactions with many subhalos would be much
more complicated and thus require a more detailed analysis.

\citet{Font01} have conducted numerical simulation of interaction between
a disk and numerous subhalos based on the CDM models. They concluded that
the effect of subhalos on a disk is rather small, and therefore subhalos
do not conflict with the presence of a thin disk since their orbit seldom take
them near the disk. However, it is worth noting that in their simulation the
initial scale height of a disk (700 pc) is already thick compared with the
observed one in the Galaxy ($\sim250$ pc), thereby leading to possibly
the underestimation of the disk heating effect. Their simulation is also
limited to only one realization of subhalos; it is yet unclear whether
the derived weak effect of subhalos on a disk is general or not.
\citet{ardi2003} have investigated more details in this disk heating
by subhalos. They found that a more massive subhalo is more effective
to heat the disk than a less massive one.
However, in their calculation subhalos are represented by rigid bodies
which never lose their mass irrespective of tidal effects of a host
galaxy, so that the disk heating is overestimated. Also, the applicability
of their result to an actual disk, especially, to what extent
accretion events of subhalos into a disk are allowed remains unclear.

Our aim in this paper is thus to set more useful limits on
the dynamical interaction between numerous subhalos and a galactic disk.
For this purpose, we conduct a series of numerical simulations, in which
a self-gravitating disk is embedded in a dark halo containing many subhalos.
In this work we set an initially thin disk with the scale height of 250 pc,
in contrast to previous numerical studies starting from the scale height of
$\sim 700$ pc much larger than the observed one (\cite{velazquez1999};
\cite{Font01}). Several models for the system of subhalos in a host halo
are taken into account in terms of their mass function, spatial distribution,
and velocity distribution. We also consider two different models for the
internal density distribution of subhalos: point-mass and extended-mass models.
In the latter model, subhalos are affected by a tidal field of a host galaxy
so that they lose their mass in the course of their orbital motions.
Based on our simulations, we investigate the dependence of the disk heating
on the model parameters and apply our analysis to understanding
an observed thin disk in the context of the disk heating by subhalos.

This paper is organized as follows. In \S~\ref{sec:model} we describe our
galaxy model which is composed of halo, bulge, and disk components. The models
of subhalos are also described in this section. In \S~\ref{sec:result} we present
the results of our numerical simulations. In \S~\ref{sec:discussion} we analyze
our results and present our prediction for the relation between the disk
heating by subhalos and an observed thin or thick disk.
Finally, in \S~\ref{sec:conclusion} we present our conclusions.

%%% Sec.2 %%%
\section{Models}
\label{sec:model}

\subsection{The Galaxy Model}

Our galaxy model is composed of three components: a disk, a bulge, and a dark
halo. To investigate the self-gravitating response of the disk component
to orbiting subhalos, we model the disk by a self-consistent N-body
realization of stars under the influence of an external force provided by
the rigid bulge and halo components. The methods of \citet{hernquist1993}
are utilized to set up the disk consisting of the distribution of N-body
particles. A detailed description of the technique can be found in
Hernquist's paper.

\begin{figure}
   \begin{center}
      \FigureFile(80mm,50mm){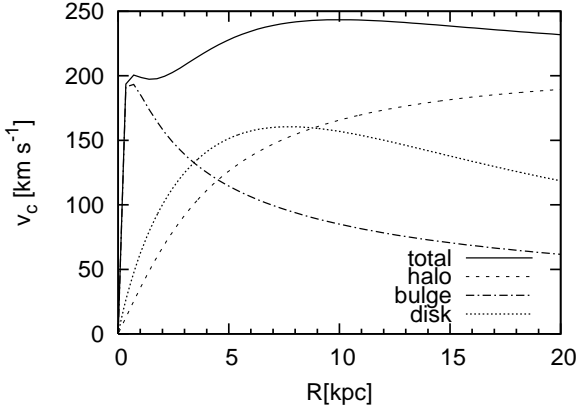}
   \end{center}
   \caption{Rotation curve for our disk galaxy model.}
 \label{fig:rotation}
\end{figure}

The density distribution of the disk is initially axisymmetric, $\rho_d(R,z)$,
using the cylindrical coordinates $(R,z)$, while the bulge and halo are
spherically symmetric, $\rho_b(r)$ and $\rho_h(r)$, respectively, using
the galactocentric distance $r$. These density distributions are given by
\begin{eqnarray}
&\rho_d&(R, z) = \frac{M_d}{4\pi R_d^2 z_d} \exp{(-R/R_d)}
        \mathrm{sech}^2{(z/z_d)} \ , \\
&\rho_b&(r)    = \frac{M_b}{2\pi}\frac{a_b}{r(a_b+r)^3} \ ,
        \label{eq:bulge} \\
&\rho_h&(r)    = \frac{M_h}{2\pi^{3/2}}
         \frac{\alpha_h}{r_c}\frac{\exp{(-r^2/r_c^2)}}{r^2+\gamma^2} \ ,
        \label{eq:halo}
\end{eqnarray}
where $M_d$, $M_b$ and $M_h$ correspond to the masses of the disk, the bulge
and the halo, respectively. The disk parameters $R_d$ and $z_d$ denote
the radial scale length and vertical scale height, respectively.
The parameter $a_b$ denotes the scale length of the bulge, while
$\gamma$ and $r_c$ are the core and cut-off radii for the halo and
$\alpha_h$ is a normalization constant. 
We choose these parameters so that the model approximately matches the
observed characteristics of the Galaxy, and the values of the parameters
are listed in Tabel \ref{table:host}. It is worth remarking that
we consider an observed thin scale height for the disk, namely $z_d = 245$ pc,
in contrast to the models by Font et al. (2001) and \citet{velazquez1999}
adopting $z_d = 700$ pc. In order to prevent the disk from gravitational
instability, we adopt a stable disk by setting Toomre's $Q$ parameter at
the solar radius $\RO = 8.5$ kpc as $Q_\odot = 1.5$. The rotation curve of
our galaxy model is shown in Figure \ref{fig:rotation}.
The rotation speed at the solar
radius is $V_c(\RO) \approx 240$ km~s$^{-1}$.

\subsection{Subhalo Models}

We construct a set of subhalo models in our numerical simulations, designated
as model A to U as tabulated in Table 2, to investigate how different physical
properties of subhalos affect the disk heating process. Each model assumption
is explained as follows.

\subsubsection{Mass function, spatial distribution and velocity anisotropy}

We consider a mass spectrum for the realization of each subhalo with a mass
$M_{\mathrm{sub}}$.
According to the results of cosmological N-body simulations by
\citet{moore1999}, \citet{klypin1999}, and \citet{ghigna2000},
this mass function can be fitted to a power law with an index of about $-2$.
We thus adopt the form,
\begin{equation}
N(M_{\mathrm{sub}})dM_{\mathrm{sub}}\propto  M_{\mathrm{sub}}^{-2} dM_{\mathrm{sub}} \ .
\label{eq:massfunc}
\end{equation}
For the convenience of numerical analysis, we set the higher and lower mass
limits for this mass function designated as $M_{\rm high}$ and $M_{\rm low}$,
respectively, and examine the role of individual subhalo masses in the
disk heating. The normalization of equation (\ref{eq:massfunc}) is given by
the total mass of the subhalo system, which is about one-tenth of
the mass of a host halo according to \citet{klypin1999} and \citet{ghigna2000}.
We thus set $0.1 M_h$ as the total mass of the subhalo system.

Recent high-resolution N-body simulations have shown that the spatial
distribution of subhalos in a host halo is less concentrated than
the host's density profile (\cite{gao2004}), which is often represented
by the so-called NFW profile (Navarro, Frenk, \& White 1997).
However, in even most recent simulations
the mass and force resolutions are yet insufficient, so the true spatial
distribution of subhalos is unclear. In this paper, instead of trying to
set a realistic spatial distribution (which is yet unknown), we adopt a
tractable model for it and attempt to extract the general results which
do not depend on this particular setting. Thus, for the initial spatial
distribution of subhalos in a host halo, we adopt
the Hernquist model \citep{hernquist1990}, in which the number density
$n(r)$ of subhalos at the galactocentric distance $r$ is given as
\begin{equation}
n(r) \propto \frac{1}{r(a+r)^3} \ ,
\end{equation}
where $a$ is the scale length in the spatial distribution.
It is worth noting that the inner density
distribution of this model is similar to that of the NFW profile.
The change of the parameter $a$ affects the incidence of subhalo-disk
interaction, which is mostly effective at $r \lesssim 10$ kpc, since smaller
$a$ yields smaller pericenters and apocenters for the orbits of subhalos.
This is highlighted in Figure \ref{fig:peri}, where the distributions of the
pericenters and apocenters of subhalos are shown for Model F ($a=87.5$ kpc)
and Model G ($a=280$ kpc), while having the same velocity distribution
(see below). It follows that the number of subhalos orbiting interior to r ~ 10 kpc 
is larger for Model F than for Model G, and the dependence of this 
number on the scale length $a$ is also seen in other models. 

For the initial velocity distribution of subhalos, we take the moments of
the collisionless Boltzmann equation following a procedure described by
\citet{hernquist1993}. The velocity ellipsoid at each spatial location is
calculated from the moment equations, and then the velocity components are
randomly selected from the Gaussian distributions for the corresponding
velocity ellipsoid. 

We adopt two different models for the velocity 
anisotropy of the subhalo system, which is parameterized by
$\beta \equiv 1-0.5(\sigma_\theta^2+\sigma_\phi^2)/\sigma_r^2$, where
$\sigma_r$, $\sigma_\theta$, and $\sigma_\phi$ are the radial, zenithal, and
azimuthal velocity dispersions, respectively.
One is the isotropic model of $\beta = 0$, which acts as
our standard model. The other is the radially anisotropic model characterized
by $\beta = 0.5$. This anisotropic model is motivated by the results of
cosmological N-body simulations (\cite{diemand2004}; \cite{abadi2006}), which
show the increase of $\beta$ with $r$, starting $\beta \sim 0$ at a halo
center to $\beta \gtsim 0.5$ in its outer parts. For the sake of simplicity,
we assume $\beta$ is constant along $r$ in our model.

\subsubsection{Effect of Baryon Condensation}

In hierarchical galaxy formation models, stars are formed in the condensation
of cooled baryon in a halo center, subsequently forming a disk component.
This condensed baryon or disk pulls the surrounding dark matter
particles inward, thereby increasing the central concentration of a dark halo
(e.g., \cite{gnedin2004}). This effect of baryon condensation is also expected
to modify the space and velocity distributions of subhalos, compared with
those obtained by dissipationless N-body simulations (\cite{gao2004}). 

We take into account this effect in our model by slowly increasing the
total masses of the disk and bulge components over a period of 10 Gyr,
after setting the initial distribution of subhalos in the presence of
a (smooth) halo alone. When the total masses of the disk and bulge
components reach the values listed in Table \ref{table:host},
the position and velocity of each subhalo are recorded for the use of
the further calculations of disk heating.
In this experiment, we treat a subhalo as a point mass and neglect the
interaction between different subhalos.

This treatment of the baryon condensation effect is admittedly highly
ideal and not self-consistent as we neglect the simultaneous modification
for a smooth halo component\footnote{Our adoption of an isothermal-like
profile for a smooth halo (equation \ref{eq:halo}), in comparison with
an NFW-like profile derived from N-body simulations, suggests
the consideration of baryon condensation for the halo setting.}.
However, the rate of the interactions between subhalos and a disk is somewhat
increased by this gravitational effect of baryonic matter, thereby
allowing us to carry out a statistically meaningful analysis on the properties
of the disk heating. In fact, this effect of baryon condensation results in
a few percent increase in the number of subhalos having pericenters smaller
than $\sim 10$ kpc, which yields the enough amount of interaction events over
the interval of numerical simulations.

\subsubsection{Internal Density Distribution}

We consider two different models for the internal density distribution of
a subhalo: point-mass and extended-mass models. In the former models,
since point-mass subhalos survive eternally in our simulations,
it is postulated that subhalos are supplied through their continuous
accretion into a host halo from outside even if some of them disappear
due to tidal destruction. In the latter models affected by tides,
we assume a King-model profile characterized by a concentration parameter
$c_{\rm King} = \log_{10}{(R_t/R_c)}$, where $R_t$ and $R_c$ denote tidal
and core radii, respectively. For these latter models, the tidal effects
of the disk (as well as the bulge and halo) on subhalos are explicitly
taken into account.
While the adoption of a King-model profile is admittedly ideal,
recent cosmological simulations by \citet{kazantzidis2004} imply that
the internal density distributions of subhalos may be described
reasonably well by a more-centrally concentrated universal profile
or the NFW profile with some tidal outer limit.
We therefore adjust our King models to match the NFW profile
in the following manner.
Firstly, based on the method outlined in NFW, we determine a set of model
parameters in the NFW profile (see Appendix \ref{appendix} for details).
Secondly, we estimate a parameter $c_{\rm King}$, whereby the half-mass
radius of the
King model is equal to that of the NFW model. Finally, we obtain a tidal
radius $R_t$ as a limiting radius of the tidal effect of a host galaxy at the
initial position of a subhalo. Thus, $R_t$ is derived from the relation
\begin{equation}
\frac{M_{\rm tot}(<r)}{r^3}=\frac{M_{\rm sub}}{R_t^3},
\end{equation}
where $M_{\rm tot}(<r)$ is the total mass of a host galaxy interior to $r$ and
$M_{\rm sub}(<R_t)$ is the mass of a subhalo. For this estimation of $R_t$, we
assume a spherically symmetric potential for a host galaxy, where the disk is
made spherical with a mass distribution $M_d(r) = M_d [1 - (1+r/R_d)
\exp(-r/R_d)]$. We also take into account the effects of baryon condensation
for getting the initial position of a subhalo. 

The parameters for point-mass models and extended-mass models that we
calculate are summarized in Table 2.

\subsection{Method for Numerical Simulation}

For the point-mass models we use a tree algorithm with a tolerance
parameter of $\theta_\mathrm{tor} = 0.7$ (\cite{barnes1986};
\cite{hernquist1987}). For the extended-mass models we use GRAPE5 systems
at the National Astronomical Observatory of Japan.
The time integration is made with the leapfrog method and a fixed
time step of 0.41 Myr. The softening length for N-body particles 
is $\epsilon = 70$ pc.
We use $N_d = 46000$ particles for the disk and the number of the subhalo
particles is listed in Table 2.
Since the mass of the subhalo is negligibly small
as compared with that of the host galaxy, we neglect the forces between
the subhalos in the point-mass models.
In contrast, for the extended-mass models, we fully take into account
the gravitational interaction between the subhalos for the convenience of
numerical calculations using GRAPE5.
We have followed the evolution up to 4.9 Gyr. 

\begin{figure}[htbp]
 \FigureFile(80mm,50mm){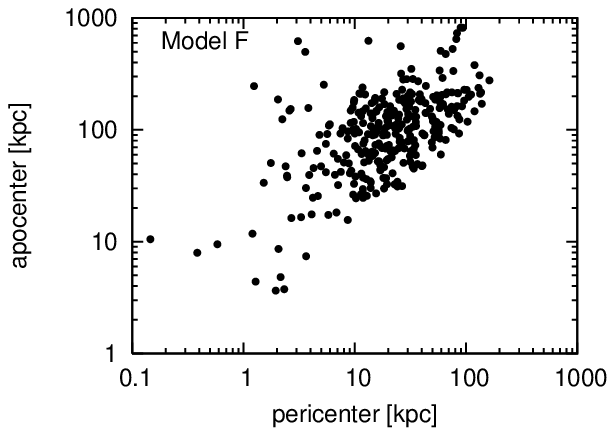}
 \FigureFile(80mm,50mm){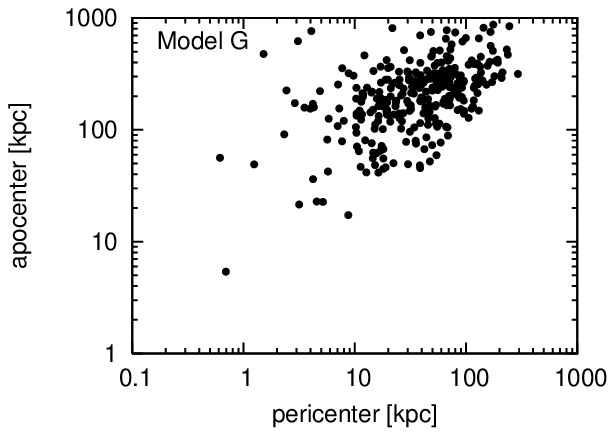}
 \caption{Distributions of the pericenter and the apocenter of the subhalos in
 model F (left panel) and model G (right panel).}
\label{fig:peri}
\end{figure}

\section{Results}
\label{sec:result}

Based on the numerical simulations of the models defined in the previous
section, we examine the effects of subhalos on the disk structure and
dynamics, especially to elucidate the dependence of several different
properties of subhalos: their internal density distribution, mass function,
and spatial and velocity distributions. 

\begin{figure*}[htbp]
 \begin{center}
  \FigureFile(65mm,50mm){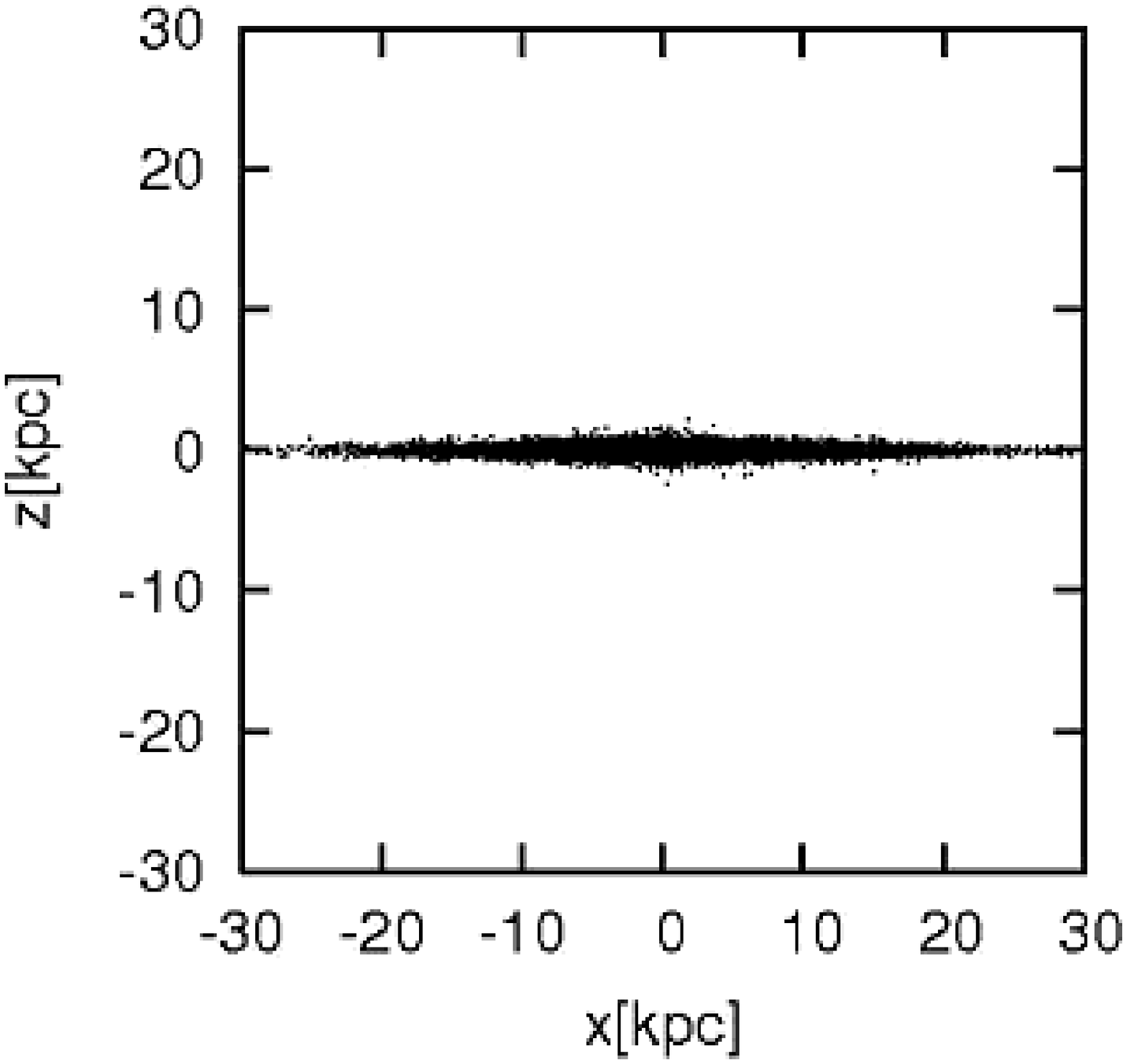}
  \FigureFile(65mm,50mm){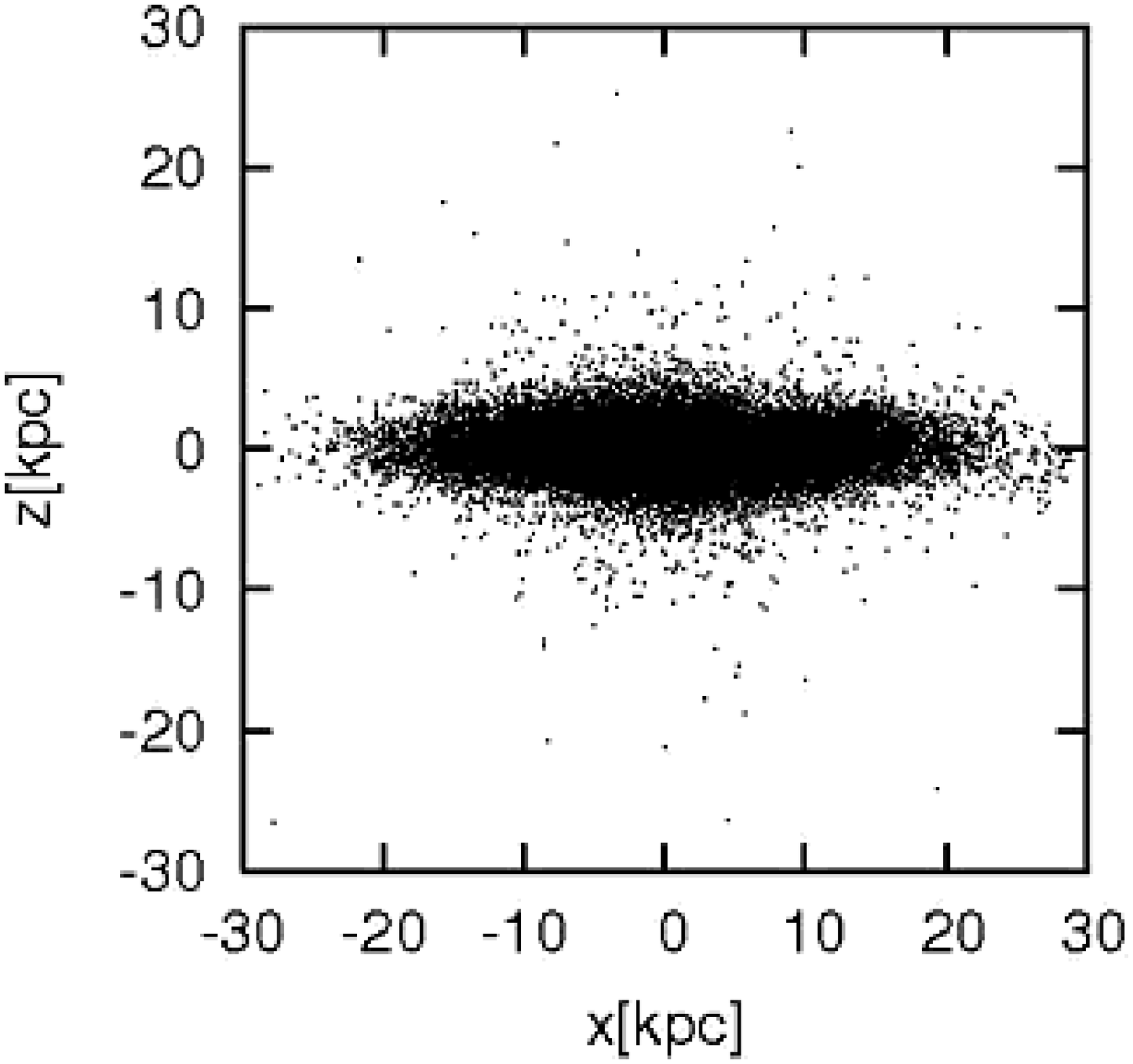}
 \end{center}
 \caption{Growth of the disk thickness for model F. The left and right panels
 show the egde-on view of the disk at the beginning ($t = 0$) and
 the end ($t = 4.9$ Gyr) of the simulation, respectively.}
\label{fig:global}
\end{figure*}

In Figure \ref{fig:global} we show the edge-on view of the disk for model F
at the beginning ($t = 0$) and the end ($t = 4.9$ Gyr) of the simulation.
The disk has been thickened and tilted by the gravitational interaction
with orbiting subhalos. To estimate the change of the disk kinematics
and thickness at specific locations, we consider the tilt of the disk
and use the axes aligned with the principal axes of the disk inertia tensor.
The heating and thickening of the disk can be
described by the changes of the velocity dispersions ($\Delta\sigma_R$,
$\Delta\sigma_z$) and by the increase of the scale height, $\Delta z_d$.
To calculate these quantities, the disk is stratified in concentric
cylindrical annuli with a width of $\delta R = 700$ pc and the particle
properties are averaged in each annulus. The scale height in each annulus 
$R$ of the disk, $z_d$, is defined by the mean square of the 
$z$-coordinates, i.e., $z_d(R) \equiv <z^2>^{1/2}$. We note that at the 
beginning of the calculations, $z_d$ is a constant of 245 pc 
as given in equation (1).

In addition to the dynamical effect of subhalos, the simulated disk is
subject to internal heating due to two-body relaxation among the disk
particles; this numerical heating always takes place in numerical
simulations with a modest number of particles.
We evaluate the effect of internal heating by evolving the disk
in isolation, i.e., in the absence of subhalos. This effect is
typically characterized
as $\Delta z_d=0.23$ kpc, $\Delta\sigma_R = 7.2$ km~s$^{-1}$,
and $\Delta\sigma_z=5.7$ km~s$^{-1}$ at the solar radius after 4.9 Gyr.
In the followings, the notation $\Delta z_d$ means the difference of
a scale height between $t = 4.9$ Gyr and $t = 0$, and
$\Delta \sigma_R$ and $\Delta \sigma_z$ also mean the change of the
velocity dispersions after 4.9 Gyr.

\subsection{Global Properties of the Disk Evolution}
\label{sec:global}

\begin{figure}[htbp]
   \begin{center}
      \FigureFile(80mm,50mm){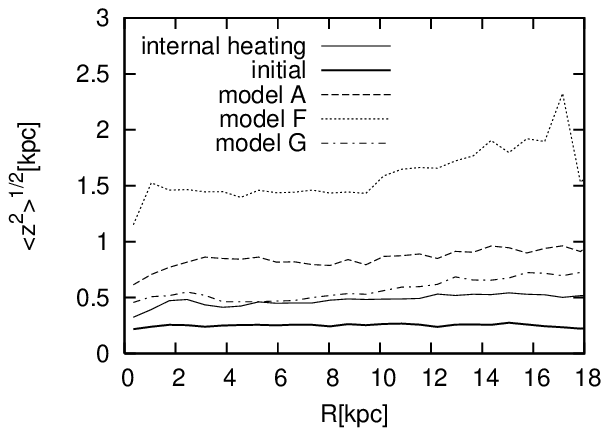}
      \FigureFile(80mm,50mm){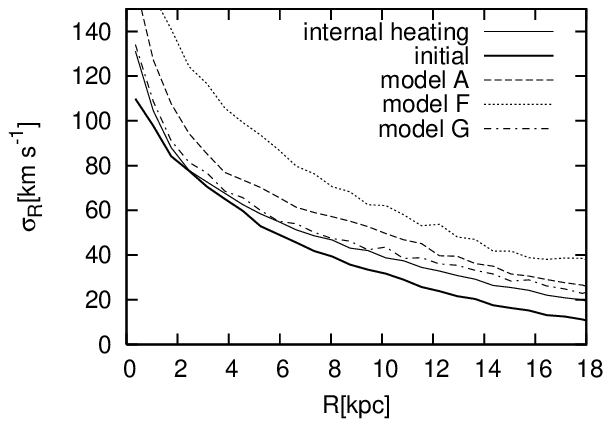}
      \FigureFile(80mm,50mm){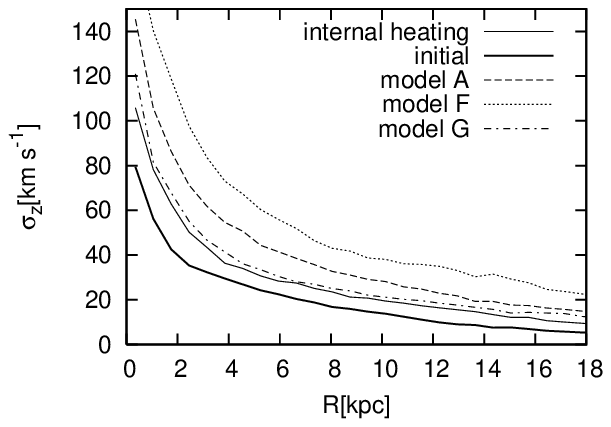}
   \end{center}
   \caption{The kinematical properties of the disk after 4.9 Gyr for
   different subhalo models. Thick solid lines show the initial setting
   and thin solid lines show the effects of internal heating in the absence
   of subhalos. Dashed, dotted, and dash-dotted lines show model A, F,
   and G, respectively.}
 \label{fig:radial-height}
\end{figure}

In Figure \ref{fig:radial-height} we show the kinematical properties of
the disk for model A, F and G, including the impact of internal heating
in our calculations.
It is evident from this figure that thickening of the disk does not
occur uniformly at all radii; given the complexity of the final disk
structure, we found it convenient to sample the kinematics at $R = \RO$,
$3R_d$, and $4R_d$, which is sufficient to provide us with a global view of
the heating and thickening.
The growth of the disk thickness in model A is $\Delta z_d \sim 0.57$ kpc,
0.61 kpc and 0.67 kpc at $R = \RO$, $3R_d$, and $4R_d$, respectively,
those values in model F are 1.19 kpc, 1.37 kpc and 1.60 kpc,
and those values in model G are 0.28 kpc, 0.31 kpc, and 0.39 kpc.
The increase in the radial and vertical velocity dispersions after 4.9 Gyr
in model A is given as $(\Delta \sigma_R, \Delta \sigma_z) = (18.8, 15.4)$,
$(17.4, 13.8)$, and $(16.6,11.9)$ km~s$^{-1}$ at $R = \RO$, $3R_d$, and $4R_d$,
respectively, those values in model F are $(32.2,26.1)$, $(30.1, 23.9)$, and
$(25.4, 22.8)$ km~s$^{-1}$, and those values in model G are $(9.9,  8.2)$,
$(10.5, 7.5)$ and $(13.5, 8.2)$ km~s$^{-1}$.

In these experiments, the main difference between model A and model F
 resides in individual masses of subhalos parameterized
by $M_{\rm high}$ and $M_{\rm low}$ (see Table 2): for model A all subhalos
have $10^8$ $M_\odot$ as $M_{\rm high} = M_{\rm low} = 10^8$ $M_\odot$,
whereas for model F the presence of more massive subhalos than
$10^8$ $M_\odot$ is allowed as $M_{\rm high} = 10^9$ $M_\odot$.
Therefore the comparison between the results of these two models highlights
the effect of individual masses of subhalos on the disk heating, where we note 
that the slight difference of the parameter $a$ between the models by a factor 
1.25 yields essentially no difference in the results. It is clear
that the presence of a few, but massive subhalos (model F) is more effective
for the disk heating than the case of many but less massive ones (model A),
as already pointed out by \citet{ardi2003}.
This suggests that the disk heating process is more sensitively enhanced
than being proportional to individual subhalo masses; massive subhalos
are more important for the disk heating.
Also, in comparison with model F, model G with the same values for
$M_{\rm high}$ and $M_{\rm low}$ yields a weak effect on the disk.
The main difference between these two models is the spatial distribution of
subhalos parameterized by $a$ ($a=87.5$ kpc and 280 kpc for model F and G,
respectively), which affects the pericenter distributions of subhalos
especially at $r \lesssim 10$ kpc (see Figure \ref{fig:peri}).
Therefore, we find that the number of subhalos which cross the disk is
also important in quantifying the disk heating.

\begin{figure}[htbp]
 \begin{center}
  \FigureFile(80mm,50mm){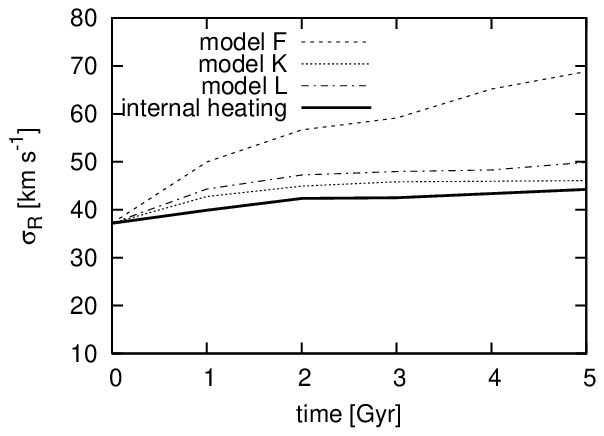}
  \FigureFile(80mm,50mm){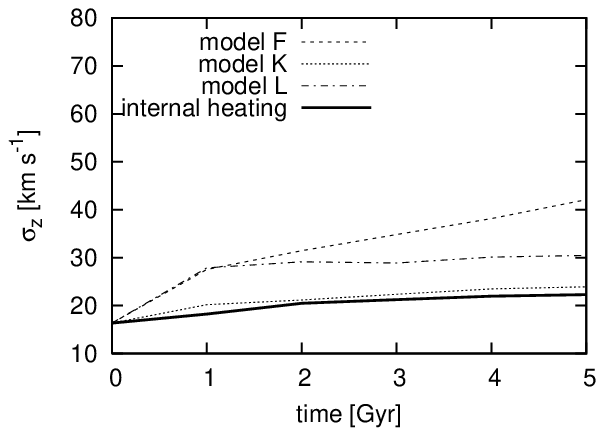}
 \end{center}
 \caption{Growth of the disk velocity dispersion in the radial and vertical
 directions at $R = \RO$ for model F (dashed lines), K (dotted lines),
 and L (dash-dotted lines). In model F subhalos are represented by point
 masses, whereas in model K and L they have a King-model profile.
 Thick solid lines show the effect of internal heating in the absence of
 subhalos.}
\label{fig:sigma}
\end{figure}

Figure \ref{fig:sigma} shows the growth of disk velocity dispersions in the
radial and vertical directions at $R=\RO$ for model F (the point-mass model),
K and L (the extended-mass model). This figure shows that the disk velocity
dispersion for model F continues to grow throughout the simulations, whereas
for model K and L the growth of the velocity dispersion almost stops
at $t \sim 1$ Gyr. It is worth noting that in these latter models subhalos
can lose their mass at the interaction with the disk, unlike the former
point-mass models in which subhalo masses remain the same.
Thus, for the extended-mass models the effect of subhalos on the disk is
temporal; subhalos appear to lose almost all of their mass at the first
interaction with the disk, so that their role in the disk heating
is effective only at the first interaction with the disk. 

\subsection{Disk Thickness vs. Subhalo Masses}
\label{section:result}

In this section we investigate in more detail the effect of orbiting
subhalos on the growth of the disk thickness.
As discussed above, the change of the disk scale height, $z_d$,
depends on $R$, in such a manner that $\Delta z_d$ is somewhat larger
at larger $R$. To quantify this disk thickening as a function of accreted
subhalo masses, we utilize the disk scale height at the outer edge
of the disk, $R = R_{\rm out}$, where the stellar light distribution
is expected to diminish steeply with $R$. This truncation of a stellar
disk has actually been observed, where the stellar light declines more
steeply than an exponential profile for a main disk and drops to low values
beyond the so-called truncation radius (van~der~Kruit \& Searle 1981a, b,
1982; \cite{kregel2002}). This usually occurs at a radius of $3-5$ disk
scale length (\cite{kregel2002}), so in our work
we adopt $R_{\rm out} = 3 R_d$ as a characteristic outer edge of the
disk, at which the change of the disk scale height is evaluated.
Also, to analyze the relation between the change of the disk
thickness and the orbits of the accreted subhalos, we estimate
the number of times which each subhalo crosses the disk region
at $R \leq 3R_d$ in the course of its orbital motion.

\begin{figure}[htbp]
 \begin{center}
  \FigureFile(80mm,50mm){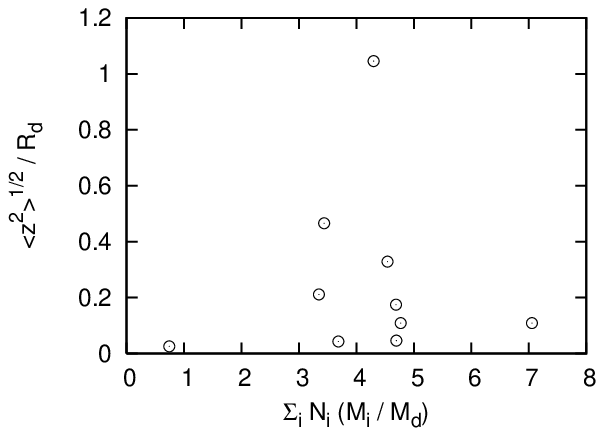}
  \FigureFile(80mm,50mm){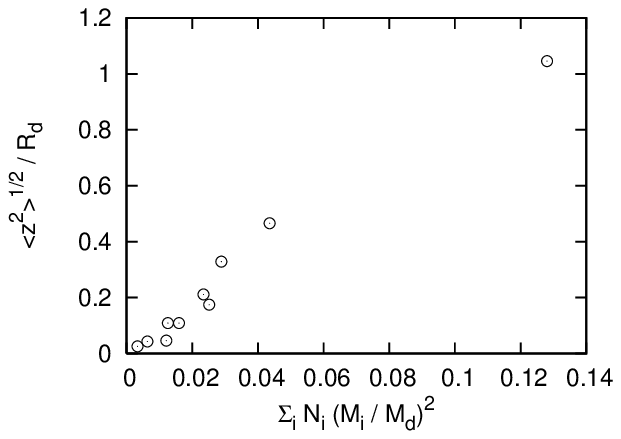}
\end{center}
 \caption{
  Relation between the growth of the disk scale height at $R = 3R_d$
  and the combination of each mass of a subhalo $M_{\mathrm{sub},i}$ and the
 number of times which it crosses the disk at $R \leq 3R_d$, denoted as
 $N_i$, for the point-mass subhalo models with $\beta=0$
 (model A to J listed in Table 2).
  The abscissas in the left and right
  panels show $\sum_i N_i (M_{\mathrm{sub},i}/M_d)$ and $\sum_i N_i
 (M_{\mathrm{sub},i}/M_d)^2$, respectively. Notice that the effect of internal
 heating has been subtracted in these plots.}
\label{fig:pt}
\end{figure}

\begin{figure}[htbp]
 \begin{center}
  \FigureFile(80mm,50mm){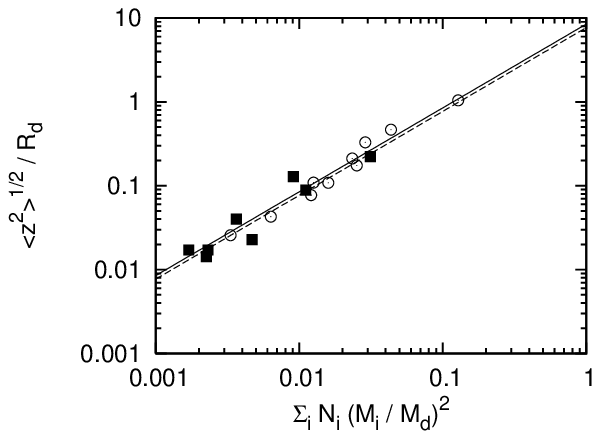}
  \FigureFile(80mm,50mm){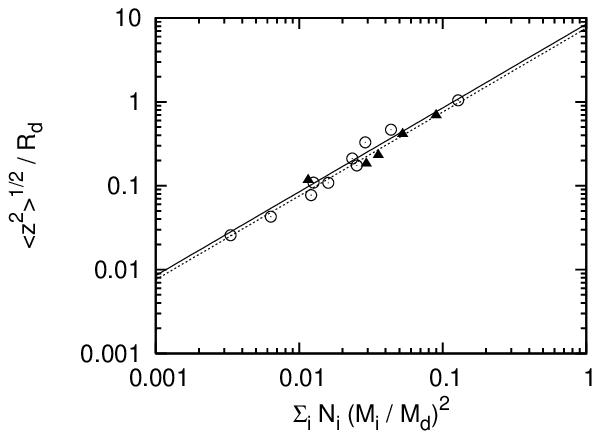}
 \end{center}
 \caption{The same as Figure \ref{fig:pt} but for several different
 subhalo models plotted in logarithmic scales.
 The left panel shows the extended-mass models with $\beta=0$ (filled squares)
 in comparison with the point-mass models with $\beta=0$ (open circles),
 whereas the right panel shows the case of an anisotropic velocity
 distribution with $\beta=0.5$ for the point-mass models (filled triangles)
 and point-mass models with $\beta=0$ (open circles).
 The solid lines show the fitting to the relations for
 the point-mass models with $\beta=0$ by the least-squares method,
 dashed lines for the extended-mass models with $\beta=0$, and dotted lines for
 the point-mass models with $\beta=0.5$.
 }
\label{fig:tidal-beta}
\end{figure}

We have carried out the simulations for several different models of
point-mass subhalos with $\beta=0$.
Figure \ref{fig:pt} shows the relation between the growth of the
disk scale height at $R = 3R_d$ (i.e. $\Delta z_d / R_d$)
and the combination of each mass of a subhalo $M_{\mathrm{sub},i}$ and the
number of times which it crosses the disk at $R \leq 3R_d$ in the course of 
its orbital motion, denoted as $N_i$
[i.e. $\sum_i N_i (M_{\mathrm{sub},i}/M_d)$ for the left panel and
$\sum_i N_i (M_{\mathrm{sub},i}/M_d)^2$ for the right panel].
Here the abscissa for
the left panel is proportional to the total accreted mass of subhalos,
whereas that for the right panel is proportional to the sum of
the squared masses of subhalos.
As is evident, while the left panel shows no correlation between the
abscissa and ordinate axes, the right panel shows a significantly tight
correlation, thereby indicating that the growth of the disk thickness
is proportional to the sum of the squared masses of subhalos;
if so, the disk thickening is more enhanced for more massive
individual subhalos as already shown in the previous subsection.
We thus investigate this relation for different subhalo models
as shown in Figure \ref{fig:tidal-beta}. In the left panel we consider
the extended-mass models with $\beta=0$ (filled squares) in comparison with
the point-mass models with $\beta=0$ (open circles).
We note here that in the extended-mass models subhalos lose almost all of their
mass at the first interaction with the disk, so we adopt $N_i=1$ for
such a case. The right panel shows the case of an anisotropic velocity
distribution with $\beta=0.5$ for the point-mass models (filled triangles)
and with $\beta=0$ for the point-mass models (open circles).
As is evident from these panels,
several different models yield an almost universal relation
between $\Delta z_d / R_d$ at $R = 3R_d$
and $\sum_i N_i (M_{\mathrm{sub},i}/M_d)^2$ at $R \leq 3R_d$.

\section{Discussion}
\label{sec:discussion}

\subsection{Dependence of the Disk Heating on Subhalo Masses}

From the results of \S~\ref{section:result}, we find that the disk thickness
is increasing with the accretion of subhalos into a disk. Our numerical
experiments suggest the following universal relation,
\begin{equation}
 \frac{\Delta z_d}{R_d} = \alpha \sum_i N_i
  \left(\frac{M_{\mathrm{sub},i}}{M_d}\right)^2 \ ,
\label{eq:relation1}
\end{equation}
where $\alpha$ is a constant of $\simeq 8$,
$R_d$ is the disk scale length, $z_d$ is the disk scale height at $R=3R_d$,
and $N_i$ is the number of times that subhalos with an individual mass of
$M_{\mathrm{sub},i}$ cross a disk
at $R \leq 3R_d$ (noting that $N_i=1$ for the extended-mass
models as subhalos lose
their mass at their first interaction with the disk).
In this expression, a subscript $i$ denotes an individual subhalo given
at $t=0$ in our simulation. An alternative, more useful expression
based on equation (\ref{eq:relation1}) is derived as follows. Supposing that
the disk has experienced the accretion of subhalos having an individual
mass of $M_{\mathrm{sub},j}$ with $j=1,...,N$ at $R \leq 3R_d$, where
the repeated accretion of a subhalo in the course of its orbital motion is
regarded as a separate accretion event with a mass
$M_{\mathrm{sub},j}$, and $N$ denotes the total number of such events.
Then, we obtain the relation,
\begin{equation}
 \frac{\Delta z_d}{R_d} \simeq 8 \sum_{j=1}^N
  \left(\frac{M_{\mathrm{sub},j}}{M_d}\right)^2 \ ,
\label{eq:relation2}
\end{equation}
which holds a more useful form for any applications than
equation (\ref{eq:relation1}). Notice that although this relation is derived
from the simulations over the interval of $4.9$ Gyr,
this is applicable to longer time evolution
by considering the total number of subhalos crossing a disk, $N$.

As equation (\ref{eq:relation2}) indicates, the increase of the disk thickness
is proportional to the square of the masses of accreted subhalos.
This mass dependency in the disk heating process can be understood if we
consider the transfer of kinetic energy from the subhalos to the disk
through dynamical friction. Here we present the summary of this
derivation and more details are shown in Appendix 2.

Firstly, the vertical equilibrium between kinetic and gravitational energy
allows us to relate the velocity dispersion $\sigma_z$ and the scale height
$z_d$ of a disk. Assuming that a disk is an isothermal sheet, we obtain
\begin{equation}
\sigma_z^2 = 2 \pi G \Sigma(R)z_d \ ,
\label{eq:height-rel}
\end{equation}
where $\Sigma(R)$ is a surface density of a disk at $R$
(e.g. \cite{spitzer1942}).
Secondly, we consider the energy input into disk stars getting through
subhalo-disk interaction: the energy loss of a subhalo is equal to
the energy pumped into a disk. This energy loss $\Delta E_\mathrm{sub}$ is
derived by the integral over an orbit of a subhalo,
\begin{equation}
 \Delta E_\mathrm{sub} = \int F_\mathrm{drag} ds \ ,
\label{eq:friction}
\end{equation}
where $F_{\mathrm{drag}}$ corresponds to a dynamical friction.
Using the Chandrasekhar formula for $F_\mathrm{drag}$, each subhalo with
a mass $M_\mathrm{sub}$ is subject to a frictional force with
$F_\mathrm{drag} \propto M_\mathrm{sub}^2$.
Finally, combining equation (\ref{eq:height-rel}) with (\ref{eq:friction}),
we obtain,
\begin{equation}
 \Delta z_d \propto \Delta \sigma_z^2 \propto \Delta E_\mathrm{sub} \sim
  F_\mathrm{drag} \cdot z_d \propto M_\mathrm{sub}^2 \ .
\end{equation}

Therefore, the dependence of $\Delta z_d$ on $M_\mathrm{sub}$ as we
have obtained in \S~3 is understood in the framework of a dynamical
friction between a subhalo and a disk.

\subsection{Comparison with an Observed Thin Disk}

Recent observations of external disk galaxies by \citet{kregel2002} have
suggested that the thickness of a (thin) disk is confined to some limiting
value relative to a scale length of a disk, which is expressed as
$z_d/R_d < 0.2$. Kregel et al. (2002) have also shown that
the distribution of $z_d/R_d$ tends to have an increasing dispersion
with increasing maximum circular velocity $V_c$ of a disk, in such a manner
that larger $V_c$ allows smaller $z_d/R_d$; conversely, a disk with
smaller $V_c$ is likely thicker.

An observed thin disk with $z_d/R_d < 0.2$ suggests that the accretion of
subhalos has been rather insignificant since a disk with a current mass
was formed.
Using equation (\ref{eq:relation2}), this observed limit implies
$(\sum_j M_{\mathrm{sub},j}^2)^{1/2} < 0.15 M_d$. Thus, we
find that an observed thin disk has not ever interacted
with subhalos with the total mass of more than 15~\% disk mass.

The dependence of $z_d/R_d$ on $V_c$ may be understood as follows.
Let $\left<M_\mathrm{sub}^2\right>$ as the mean square of a subhalo
mass, defined as $\left<M_\mathrm{sub}^2\right>
= \sum_j^N M_{\mathrm{sub},j}^2 / N$. Then, 
equation (\ref{eq:relation2}) can be written as
\begin{equation}
 \frac{\Delta z_d}{R_d} \propto
 N \frac{\left<M_\mathrm{sub}^2\right>}{M_d^2} \ .
\label{eq:result1}
\end{equation}
We suppose that $N$, the number of subhalos which cross a disk at
$R \le 3R_d$, is roughly proportional to the spherical volume with a radius
$3R_d$ inside a dark halo, given the spatial distribution of subhalos.
This reads $N \propto R_d^3$. Also suppose that the central disk surface
density $\Sigma_0 = M_d/(2\pi R_d^2)$ is nearly the same for a disk
galaxy, which may correspond to nearly the same central surface
brightness for a bright disk (e.g., \cite{freeman1970}).
Then equation (\ref{eq:result1}) is rewritten as,
\begin{equation}
 \frac{\Delta z_d}{R_d}\propto 
  \left<M_\mathrm{sub}^2\right> \cdot V_c^{-2} \ ,
\label{eq:result2}
\end{equation}
where $V_c$ is a disk circular velocity derived from
$V_c \propto GM_d / R_d$.

Thus, if the observed disk thickness is controlled by disk-subhalo
interaction, i.e., $z_d \sim \Delta z_d$, and 
$\left<M_\mathrm{sub}^2\right>$ is roughly the same for each disk galaxy,
then we find that the effect of subhalos on a disk with a smaller mass
is more significant than a disk with a larger mass.
This is in good agreement with the observation (\cite{kregel2002})
that a disk with smaller $V_c$ is likely thicker.

\subsection{The Relation to the Origin of a Thick Disk}

In recent years, new datasets for the thick disk component of the Galaxy
as well as for a thick disk of an external galaxy have been available,
showing several important properties of a thick disk.
Based on the third data release of the Sloan Digital Sky Survey
(\cite{york2000}), \citet{allende2006} have found that the stars belonging
to the thick disk have no vertical metallicity gradient. The rotational
velocity for the same stars however shows a vertical gradient of
$\sim -16$ km~s$^{-1}$~kpc$^{-1}$ between 1 and 3 kpc from the Galactic
plane. The observations of an external disk galaxy (\cite{yoachim2006})
have shown that the ratio of the total luminosity of a thick disk
to that of a thin one is related to the circular velocity of a galaxy,
in such a manner that the ratio tends to increase with decreasing
circular velocity; a less massive galaxy has a brighter (and possibly
more massive) thick disk relative to a thin one. These properties of
a thick disk are expected to set important constraints on its origin.

Here, we consider a possibility that a thick disk has been formed
by the dynamical effect of numerous dark subhalos on a pre-existing disk.
This interaction effect may be rather weak at the current epoch
as only a small fraction of subhalos would cross a disk (\cite{Font01}),
but the effect at early times may have been more significant due to a smaller
disk mass and larger accretion rate of subhalos onto a parent halo.
If so, a pre-existing disk should have suffered from considerable heating
at early times, which results in the formation of a thick disk; subsequent
slow accumulation of baryonic gas in a plane may form a thin disk component.

\begin{figure}[htbp]
 \begin{center}
  \FigureFile(65mm,50mm){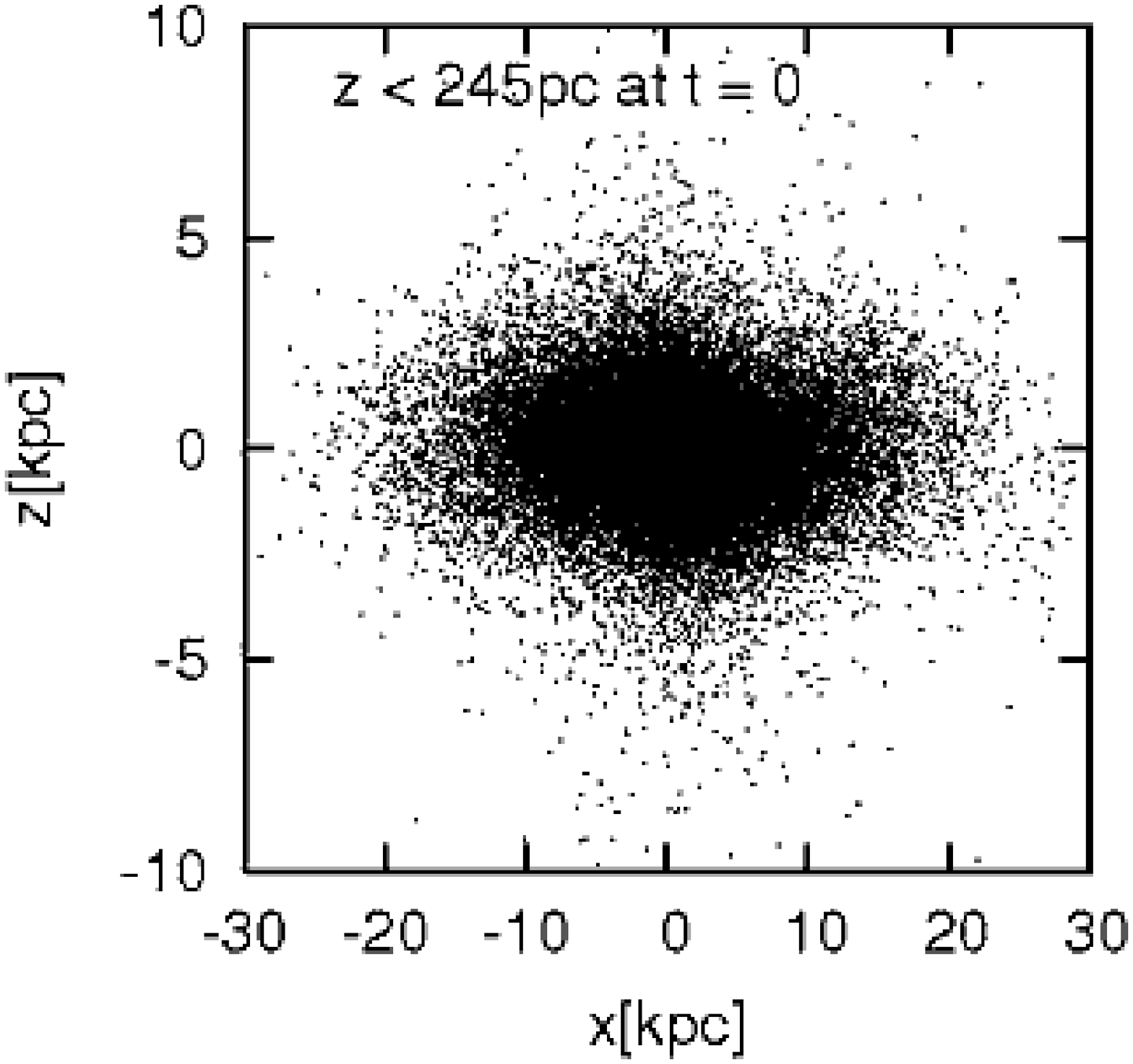}
  \FigureFile(65mm,50mm){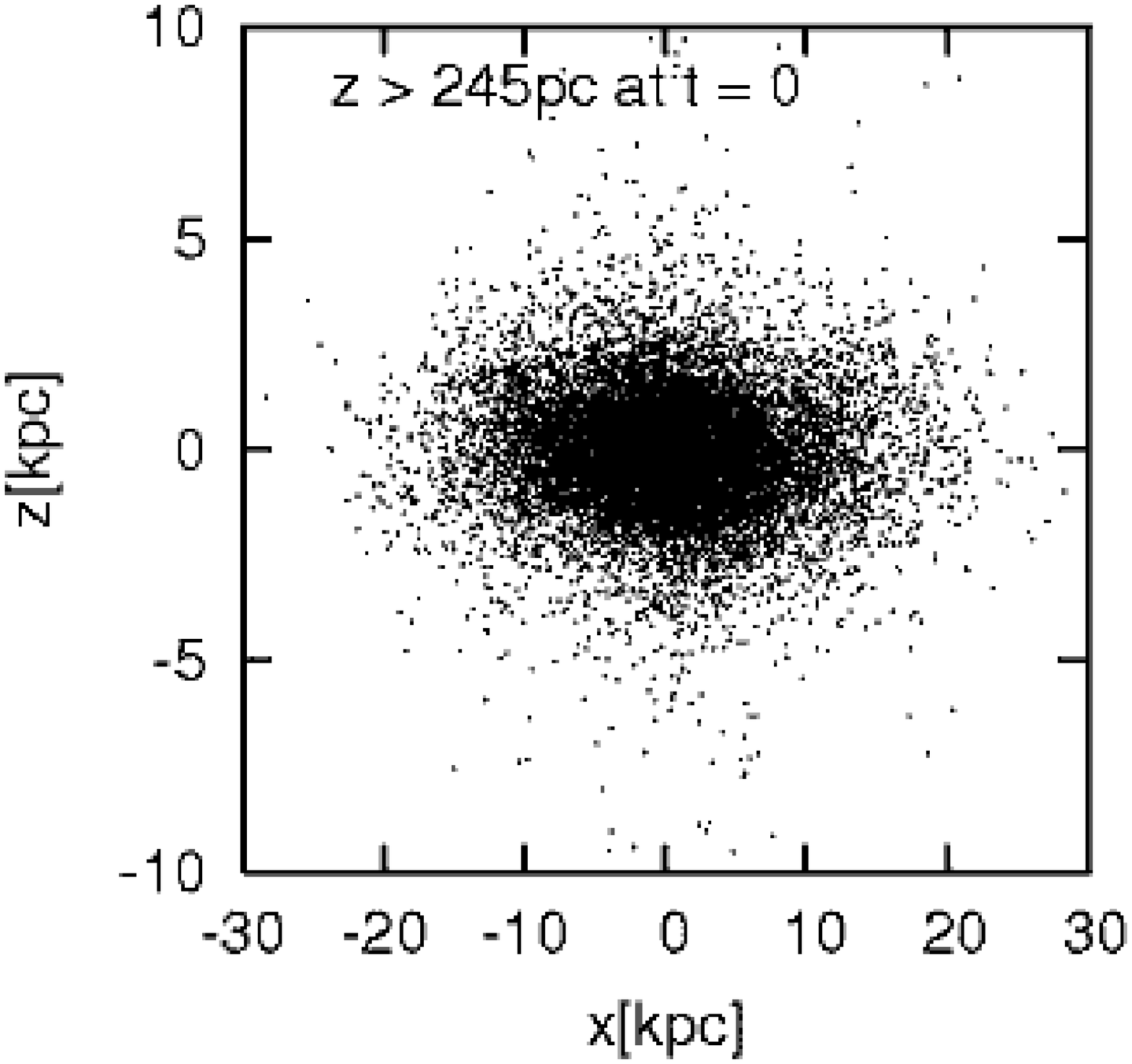}
 \end{center}
 \caption{Distribution of stars in model F at the end of the simulation
 ($t=4.9$ Gyr), differentiated by their initial ($t=0$) positions
 at $z<245$ pc (left panel) and $z>245$ pc (right panel).}
\label{fig:thick}
\end{figure}

\begin{figure}[htbp]
 \begin{center}
  \FigureFile(80mm,50mm){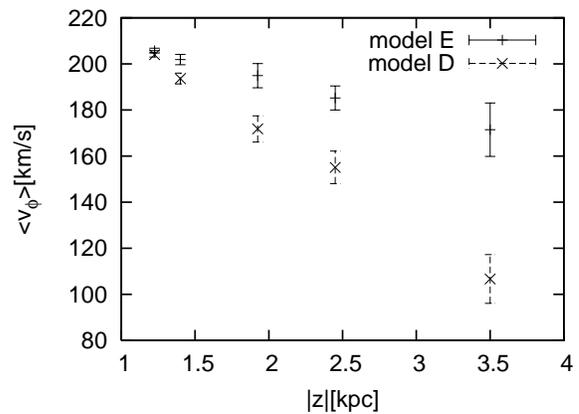}
 \end{center}
 \caption{Mean azimuthal velocity of the stars located at annulus
  $R=\RO \pm 1.05$ kpc as a function of the distance away from the plane,
  for model D and E.
  %The particle properties are averaged
  %bin with a width of $350$ pc, $350$ pc, $350$ pc, $700$ pc, and $1.4$ kpc
  %starting from $z=1.05$ kpc.
  Note that both models show similar vertical scale
  height $z_d \approx 1$ kpc at the end of the simulations ($t = 4.9$ Gyr).}
 \label{fig:vc-grad}
\end{figure}

To assess this possibility in light of the observed properties of
a thick disk, we investigate the properties of the disk thickened by
subhalos in our model. Figure \ref{fig:thick} shows the distribution of the
stars in model F at the end of the simulation ($t=4.9$ Gyr), differentiated
by the initial ($t=0$) positions at either $z<245$ pc (left panel)
or $z>245$ pc (right panel). As is evident, the disk stars are well mixed
by the disk heating. This suggests that even if a pre-existing disk had
a metallicity gradient, the disk heating by interactions with subhalos
would have wiped out this gradient, leaving a thick disk in agreement with
the observations. It is also suggested that this disk heating process
prompts a vertical gradient in rotational velocity. This is explained as
follows. A heated disk puffs up not only in the vertical but also in the
radial direction, where the stars located at high $|z|$ have large radial
velocity dispersion compared with those at low $|z|$ and thus have small
rotational velocities on average due to the effect of asymmetric drift:
given a gravitational force inward, the increase of radial pressure of stars
reduces the effect of a centrifugal force. This vertical gradient in
rotational velocity is actually obtained in our simulation models as shown
in Figure \ref{fig:vc-grad}. It is found that the gradient amounts to
$-10 \sim -30$ km~s$^{-1}$~kpc$^{-1}$ between 1 and 3 kpc from the plane,
in good agreement with the observations.
This experiment suggests that the presence of a
vertical gradient in rotational velocity of a thick disk may be an
important clue to distinguishing the scenarios for the origin of a thick disk;
models invoking monolithic disk collapse (\cite{burkert1992}) or chaotic
merging event of building blocks (\cite{brook2004}) at early times of a
galaxy may have difficulties in this regard.

More definite picture for the formation of a thick disk must await more
elaborate modeling of a forming galaxy in the context of hierarchical
clustering. In particular, each galaxy has a different merging history
of subhalos and so different spatial and velocity distributions, which
inevitably affects the interpretation for the observed properties of
a thick disk, such as for its fraction as a function of a disk mass.
Also, the observations show that almost all of disk galaxies have a
distinct thick disk in addition to a thin component. This implies that
the dynamical effects of subhalos on a disk were significant only before
a specific epoch and the subsequent formation of a thin disk has been
unaffected by subhalos, although it is yet unclear if it is applicable to
all galaxy-sized halos having different merging histories.
Therefore, to assess the scenario, more detailed numerical studies are
required, taking into account the growth of both a dark halo and
disk simultaneously.

\section{Conclusions}
\label{sec:conclusion}

We summarize our conclusions as follows:

\begin{itemize}
 \item The dynamical effects of subhalos on a disk are represented by
  the relation between the change of the disk scale height $\Delta z_d$
  (measured at the disk edge $R=3R_d$) and individual masses of subhalos
  $M_{\mathrm{sub}}$, i.e.,
  $\Delta z_d/R_d \simeq 8 \sum_{j=1}^N (M_{\mathrm{sub},j}/M_d)^2$,
  where $R_d$ is a disk scale length, $M_d$ is a disk mass, and
  $N$ is the total number of accretion events of subhalos inside a 
  disk region ($\le 3R_d$). 
 \item If subhalos with the total mass of more than 15~\% disk mass
  interact with a disk, then the disk thickness is made larger than
  the observed range.
 \item A less massive disk with smaller circular velocity $V_c$ is found to
  be more affected by subhalos than a disk with larger $V_c$, which is
  in agreement with the observed properties of a thin disk.
 \item Stars in a significantly thickened disk by subhalos appear to be
  well mixed and show a vertical gradient in their rotation velocity,
  being similar to the observed properties of the thick disk in the Galaxy.
\end{itemize}

We note that the relation (\ref{eq:relation2}) we have obtained here is
universal and thus useful for the applications to any relevant issues,
including the dynamics of an evolving stellar disk at the center of
a growing dark halo. Such detailed studies of a galactic disk in
comparison with recently increasing datasets of a remote disk galaxy will be
of great importance and is left to future work.

\bigskip
The numerical computations reported here were carried out on GRAPE systems
(project ID: g05b05) kindly made available by
the Astronomical Data Analysis Center (ADAC) at the National Astronomical
Observatory of Japan (NAOJ).
This work has been supported in part by a Grant-in-Aid for Scientific Research
(15540241, 17540210) from the Japanese Ministry of Education, Culture, Sports,
Science and Technology.

\appendix
%%% Appendix 1 %%%
\section{NFW profile}
\label{appendix}
The NFW density profile is given by 
\begin{equation}
 \rho = \rho_{\mathrm{crit}} \frac{\delta_0}{(r/r_s)(1+r/r_s)^2} \ ,
\end{equation}
where $\rho_{\mathrm{cirt}}$ is the critical density of the Universe,
$r_s$ is a scale radius, and $\delta_0$ is a characteristic density contrast.
Following NFW, we define the limiting radius of a virialized halo,
$r_{\mathrm{200}}$, to be the radius within which the mean mass density is
$200\rho_{\mathrm{crit}}$. 
Also, the concentration parameter of a halo is defined as
$c=r_{\mathrm{200}}/r_s$, with which $\delta_0$ is given as
\begin{equation}
 \delta_0=\frac{200}{3}\frac{c^3}{[\ln(1+c)-c/(1+c)]} \ .
\label{eq:delta-c}
\end{equation}

To put the NFW density profile in a cosmological context, we need to calculate
the concentration factor $c$, which is related to $\delta_0$
via equation (\ref{eq:delta-c}). The appropriate value of $c$ depends on
halo formation history and on cosmology. NFW proposed a simple model for $c$
based on halo formation time. The formation redshift $z_{\mathrm{coll}}$ of
a halo identified at $z = 0$ with mass $M$ is defined as the redshift
by which half of its mass is in progenitors with mass exceeding $fM$,
where $f$ is a constant. With this definition, $z_{\mathrm{coll}}$
can be computed by simply using the Press-Schechter formalism
(e.g. \cite{lacey1993}),
\begin{equation}
 \mathrm{erfc}
  \left\{
   \frac{\delta_{\mathrm{c}}(z_{\mathrm{coll}})-\delta_\mathrm{c}^0}
   {\sqrt{2[\Delta_0^2(fM)-\Delta_0^2(M)]}}
  \right\}
  = \frac{1}{2} \ ,
\label{eq:formation}
\end{equation}
where $\Delta_0^2(M)$ is the linear variance of the power spectrum at $z=0$
smoothed with a top-hat filter of mass $M$, $\delta_{\mathrm{c}}(z)$ is the
density threshold for spherical collapse by redshift $z$, and
$\delta_{\mathrm{c}}^0=\delta_{\mathrm{c}}(0)$. NFW found that the
characteristic overdensity of a halo at $z=0$ is related to its formation
redshift $z_{\mathrm{coll}}$ by
\begin{equation}
 \delta_0(M, f) = C(f) \Omega_0[1+z_{\mathrm{coll}}(M,f)]^2 \ ,
\label{eq:delta0}
\end{equation}
where the normalization $C(f)$ depends on $f$ and $\Omega_0$ is
the current density parameter of the Universe. We will take $f=0.01$ as
suggested by the N-body results of NFW. In this case 
$C(f)\approx 3\times 10^3$. Thus, for a halo of given mass at $z=0$, one can
obtain the concentration factor $c$ from
equations (\ref{eq:delta-c})-(\ref{eq:delta0}). In practice, we first solve
$z_{\mathrm{coll}}$ from equation (\ref{eq:formation}) and insert the value of
$z_{\mathrm{coll}}$ into equation (\ref{eq:delta0}) to get $\delta_0$. We
then use this value of $\delta_0$ in equation (\ref{eq:delta-c})
to solve for $c$.

In this experiment, we adopt a standard set of cosmological parameters
as $\Omega_0=0.3$, $\Lambda =0.7$, $h=0.7$, and $\sigma_8=1.3$,
where $\Lambda$ is a cosmological constant, $h$ is a normalized Hubble
constant of $h \equiv H_0 / 100$~km~s$^{-1}$~Mpc$^{-1}$, and $\sigma_8$
parameterizes density fluctuations at $8~h^{-1}$ Mpc.

%%% Appendix 2 %%%
\section{Derivation of equation (\ref{eq:relation2})}

We show here the derivation of equation (\ref{eq:relation2}) based on
the following dimensional analysis.

Firstly, we derive the relation between the vertical velocity dispersion
$\sigma_z$ of disk stars and the scale height $z_d$ of a self-gravitating
axisymmetric disk in cylindrical coordinates $(R,z)$.
For the limit of a thin disk where $z$-derivatives dominate over $R$-derivatives,
an equation for vertical equilibrium and a Poisson equation read, respectively,
\begin{equation}
 \frac{1}{\rho_d}\frac{\partial (\rho_d \sigma_z^2)}{\partial z}
  + \frac{\partial \Phi}{\partial z} \simeq 0 \ ,
 \quad
 \frac{\partial^2 \Phi}{\partial z^2} \simeq 4 \pi G \rho \ ,
\end{equation}
where $\rho_d$ is a mass density of a disk and $\Phi$ is a gravitational
potential. Integrating the second equation over all $z$ 
gives the surface mass density $\Sigma(R)$ vs. $\Phi$, i.e.,
$2 \pi G \Sigma(R) \simeq \partial \Phi / \partial z$. Then, inserting this
into the first equation, and assuming $\rho_d \propto \exp(-z/z_d)$ and
$\sigma_z$ is constant, we obtain
$z_d \simeq \sigma_z^2 / 2 \pi G \Sigma(R)$.
This equation suggests that the change of the disk thickness is
related to the change of the velocity dispersion, i.e.,
\begin{equation}
\Delta z_d \simeq \frac{\Delta \sigma_z^2}{2 \pi G \Sigma(R)} \ .
\label{eq:app-height-rel}
\end{equation}

Secondly, the change of the velocity dispersion $\Delta \sigma_z^2$
is related to the energy input into disk stars getting through
subhalo-disk interaction: the energy loss of a subhalo is equal to
the energy pumped into a disk. Denoting this energy loss as
$\Delta E_\mathrm{sub}$ per a subhalo, the resultant
$\Delta \sigma_z^2$ for a disk with a total mass $M_d$ reads,
\begin{equation}
\Delta \sigma_z^2 = \frac{2 \Delta E_\mathrm{sub}}{M_d} \ .
\end{equation}
We note that $\Delta E_\mathrm{sub}$ is derived by the integral over
an orbit of a subhalo,
\begin{equation}
 \Delta E_\mathrm{sub} = \int F_\mathrm{drag} ds \sim F_\mathrm{drag}z_d \ ,
\label{eq:app-friction}
\end{equation}
where $F_{\mathrm{drag}}$ corresponds to a dynamical friction.
Using the Chandrasekhar formula for $F_\mathrm{drag}$, each subhalo with
a mass $M_\mathrm{sub}$ is subject to a frictional force with
\begin{equation}
  F_\mathrm{drag} = \frac{4\pi G^2 M_\mathrm{sub}^2 \rho_d
  \ln \Lambda}{v_\mathrm{sub}^2}
  \left[\mathrm{erf}(X)-\frac{2X}{\sqrt{\pi}}e^{-X^2}\right] \ ,
\end{equation}
where $\ln \Lambda$ is the Coulomb logarithm, 
$X\equiv v_\mathrm{sub}/\sqrt{2}\sigma$ with $\sigma$ being the disk velocity
dispersion, and $v_\mathrm{sub}$ is the subhalo's velocity.
For $v_\mathrm{sub}$, we suppose that it is represented by a virial velocity
of a smooth dark halo with a mass $M_H$ and a radius $r_H$, giving
$v_\mathrm{sub}^2 \sim G M_H / r_H$. Furthermore, if $M_d$ and $R_d$ for
a disk is some fraction of $M_H$ and $r_H$, given as $M_d = f_1 M_H$ and
$R_d = f_2 r_H$, where typically $f_1 \sim f_2 \sim O(10^{-1})$,
we obtain $v_\mathrm{sub}^2 \sim G M_d / R_d$.
For $\rho_d$, we set $\rho_d \sim M_d / R_d^2 z_d$.

Finally, using the above equations,
the change of the disk thickness induced by $N$ accretion events of
subhalos is estimated by,
\begin{eqnarray}
\Delta z_d &\propto& \frac{N \Delta \sigma_z^2}{G\Sigma} 
 \propto \frac{N G^2 z_d M_\mathrm{sub}^2 \rho_d /v_\mathrm{sub}^2}{G\Sigma M_d}
            \nonumber \\
&\propto& \frac{N M_\mathrm{sub}^2}{\Sigma M_d R_d} \ .
\end{eqnarray}
Thus, we obtain
\begin{equation}
\frac{\Delta z_d}{R_d} \propto N \frac{M_\mathrm{sub}^2}{M_d^2} \ ,
\end{equation}
which is consistent with equation (\ref{eq:relation2}).

\newpage

\begin{table}
\begin{center}
\caption{Galactic parameters}
\label{table:host}
\begin{tabular}{ccc} \hline\hline
 & Symbol  & Value \\\hline
Disk: &  &  \\
 & $N_d$\footnotemark[$*$]& $46000$ \\
 & $M_d$ & $5.6 \times 10^{10}$ $\MO$ \\
 & $R_d$ & $3.5$ kpc \\
 & $z_d$ & $245$ pc  \\
 & $Q_\odot$ & $1.5$ \\
 & $\RO$ & $8.5$ kpc \\
 & $\epsilon$ & $70$ pc \\\hline
Bulge: &  &  \\
 & $M_b$ & $1.87 \times 10^{10}$ $\MO$ \\
 & $a_b$ & $525$ pc \\\hline
Halo: &  &  \\
 & $M_h$ & $7.84 \times 10^{11}$ $\MO$ \\
 & $\gamma$ & $3.5$ kpc \\
 & $r_c$ & $84$ kpc \\\hline
\multicolumn{3}{@{}l@{}}{\hbox to 0pt{\parbox{85mm}{\footnotesize
       \footnotemark[$*$] The number of particles used for the disk.
     }\hss}}
\end{tabular}
\end{center}
\end{table}

\newpage

\begin{table}[htbp]
\begin{center}
\caption{The parameters of the subhalos}
\label{}
\begin{tabular}{c c c c c c} \hline\hline
Model & Number of & $a$ & $M_\mathrm{high}$ & $M_\mathrm{low}$ & 
      $n_\mathrm{sub}$\footnotemark[$*$]\\
      & subhalos  &[$\mathrm{kpc}$] & [$M_\odot$]    & [$M_\odot$] & \\\hline
\multicolumn{5}{c}{point-mass models with $\beta=0$} \\ \hline\hline
A & $784$ & $70$ & $10^8$ & $10^8$ &  \\
B & $784$ & $140$ & $10^8$ & $10^8$ &  \\
C & $392$ & $140$ & $2 \times 10^8$ & $2 \times 10^8$ &  \\
D & $261$ & $140$ & $3 \times 10^8$ & $3 \times 10^8$ &  \\
E & $200$ & $175$ & $4 \times 10^8$ & $4 \times 10^8$ &  \\
F & $318$ & $87.5$ & $10^9$ & $10^8$ &  \\
G & $313$ & $280$ & $10^9$ & $10^8$ &  \\
H & $175$ & $140$ & $10^{10}$ & $10^8$ &  \\
I & $1141$ & $175$ & $10^{10}$ & $10^7$ &  \\
J & $1959$ & $140$ & $10^9$ & $10^7$ &  \\\hline
\multicolumn{6}{c}{extended-mass models with $\beta=0$} \\ \hline\hline
K & $318$ & $87.5$ & $10^9$ & $10^8$ & 182 \\
L & $175$ & $140$ & $10^{10}$ & $10^8$ & 182 \\
M & $362$ & $24.5$ & $10^9$ & $10^8$ & 182 \\
N & $200$ & $175$ &  $4 \times 10^8$ &  $4 \times 10^8$ & 182 \\
O & $112$ & $70$ & $7 \times 10^8$ & $7 \times 10^8$ & 182 \\
P\footnotemark[$\dagger$] & $280$ & $52.5$ & $10^{10}$ & $10^8$ & 170 \\
Q & $173$ & $24.5$ & $10^{10}$ & $10^8$ & 170 \\\hline
\multicolumn{5}{c}{point-mass models with $\beta=0.5$} \\ \hline\hline
R & $197$ & $140$ & $10^{10}$ & $10^8$ &  \\
S & $362$ & $87.5$ & $10^9$ & $10^8$ &  \\
T & $249$ & $140$ & $3\times 10^{9}$ & $10^8$ &  \\
U & $361$ & $157.5$ & $10^9$ & $10^8$ &  \\\hline
\multicolumn{6}{@{}l@{}}{\hbox to 0pt{\parbox{85mm}{\footnotesize
       \par\noindent
       \footnotemark[$*$] The number of particles used for each subhalo.
       \par\noindent
       \footnotemark[$\dagger$]
       Only in this model, the total mass of the subhalo system is 13~\% of
       that of the host galaxy.
     }\hss}}
\end{tabular}
\end{center}
\end{table}

\end{document}